\documentclass[doublecol,figures]{epl2} 
\usepackage{amssymb,amsmath}
\newcommand{\bp}{{\bf p}}
\newcommand{\bq}{{\bf q}}
\newcommand{\bk}{{\bf k}}

\newcommand{\bP}{{\bf P}}
\newcommand{\beq}{\begin{equation}}
\newcommand{\eeq}{\end{equation}}

\title{Polarons and dressed molecules near narrow Feshbach resonances}
\shorttitle{Polarons and dressed molecules near narrow Feshbach resonances}

\author{P. Massignan}
\shortauthor{P. Massignan}
\institute{ICFO - The Institute of Photonic Sciences - 08860 Castelldefels (Barcelona), Spain}


\abstract{The properties of impurities immersed in a large Fermi sea are naturally described in terms of dressed quasiparticles: attractive and repulsive polarons, and dressed molecules. Motivated by recent experiments on narrow Feshbach resonances, we analyze here how the quasiparticle properties are affected by a non-zero resonance range. We find two interesting analytic results. For large range, the ground state energy close to resonance is shown to become perturbative in the inverse range. In the limit of broad resonance instead, we provide a new Tan's relation linking the impurity ground state energy $E_\downarrow$ to the number of atoms in its dressing cloud $\Delta N$. As a corollary, at unitarity one finds $\Delta N=-E_\downarrow/\epsilon_F $, with $\epsilon_F$ the Fermi energy of the bath.}

\begin{document}

\maketitle
\date{\today}
\section{Introduction}
The study of the static and dynamic properties of impurities in liquids and crystals provided us with a wealth of information on various condensed matter systems. As a celebrated example, we may recall how Landau showed that the strong interactions experienced by electrons propagating in a crystal could be described in very intuitive terms by introducing dressed quasiparticles named polarons \cite{Landau33}. The investigation of magnetic impurities brought to the discovery of the Kondo effect \cite{Kondo64}, fundamental to understand the electric resistivity of many materials at low temperature, while charged impurities in Helium liquids \cite{Schwarz07} played a key role in evidencing vortex lattices in the superfluid.

Ultracold gases provide an optimal framework for studies of many-body quantum effects \cite{Bloch08,Giorgini08}. The first realization of polarons in an imbalanced homo-nuclear fermionic mixture was reported in an enlightening experiment \cite{Schirotzek09}, where the energy and residue of the attractive polaron were measured via radio-frequency (rf) spectroscopy. This and later experiments \cite{Schirotzek09,Nascimbene10} confirmed that  the dressed quasiparticles interact very weakly with each other.
The polaron effective mass was investigated by means of collective oscillations in ref.~\cite{Nascimbene09}, and recently the complete spectral response of the system was mapped out in ref.~\cite{Kohstall11}. There, by exploiting a novel rf spectroscopy scheme, the energy of attractive and repulsive polarons was measured, and a broad molecular continuum detected. The coherent nature of the polarons was probed by driving Rabi oscillations, providing a precise measurement of their residue.

The great majority of measurements on strongly-interacting fermions employed broad Feshbach resonances, but narrow resonances are becoming of increasing relevance as a new generation of experiments started investigating hetero-nuclear fermionic mixtures \cite{Wille08,Tiecke10,Trenkwalder11}, where most known resonances are coincidentally narrow \cite{Chin10}.  From a theoretical point of view, at broad resonances the relevant physics is a function of a single parameter, the scattering length $a$. The main aim of this paper is the study of the richer physics of narrow resonances, where quantities depend strongly on a second parameter, the resonance range $R^*$.

We wish to investigate the low-energy excitation spectrum of a single impurity $\downarrow$ immersed in a large Fermi gas of particles $\uparrow$. In presence of a broad resonance, the physical scenario is by now well understood \cite{Chevy06,Combescot07,Prokofev08,Combescot09,Mora09,Punk09,Cui10,Mathy11,Massignan11,Schmidt11}.
On the BCS side, the impurity experiences a weak attractive interaction with the Fermi sea, yielding a thinly dressed quasiparticle, the attractive polaron. As the attraction grows beyond a critical value, it becomes energetically favorable for the impurity to bind tightly to an atom in the gas, thereby forming a dressed molecule. When the impurity is very light ($m_\uparrow/m_\downarrow\gtrsim 7$), it becomes possible to form even trimers \cite{Mathy11}. At positive energies, there exists a second branch of polarons, which experience repulsive interactions with the gas and are intrinsically metastable, due to the presence of lower-lying excitations.

To be able to investigate narrow resonances, we employ in this paper a many-body two-channel model. In the strong coupling (broad resonance) limit, our model reduces to earlier single-channel approaches, and there it has been validated by an ab-initio Monte-Carlo calculation \cite{Prokofev08}. Since we extend earlier treatments towards a weak coupling regime (narrow resonance) where the model becomes perturbative in the inverse range and exactly solvable \cite{Gurarie07}, we expect the results presented here to be accurate throughout.

\section{Many-body scattering at narrow resonances}
The key ingredient in many-body scattering at low energies is the T-matrix $T(\bP,\omega)$, describing the collision of an $\uparrow$ and a $\downarrow$ atom, with masses $m_{\uparrow}$ and $m_{\downarrow}$, reduced mass $m_r$, total momentum $\bP$ and energy $\omega$ (we have set $\hbar=1$).

 At low energies higher partial waves play a negligible role, and we may safely restrict ourselves to $s$-wave scattering only, parametrized by the scattering length $a$.
The usual single-channel T-matrix then reads 
\beq
T_{\rm 1ch}^{-1}(\bP,\omega)=m_r/(2\pi a)-\Pi(\bP,\omega).
\eeq
 The propagation of the two particles in the medium between scattering events is described by the renormalized pair propagator $\Pi(\bP,\omega)$
\beq
\Pi(\bP,\omega)=\int\frac{{\rm d}\bk}{(2\pi)^3}\left[\frac{1-f_\uparrow(\bk)-f_\downarrow(\bP+\bk)}{\omega+i0^{+}-\xi_{\uparrow\bk}-\xi_{\downarrow\bP+\bk}}+\frac{2m_r}{k^2}\right],
\label{pairPropagator}
\eeq
with $f_\sigma(\bk)=[\exp(\beta\xi_{\sigma\bk})+1]^{-1}$ the Fermi function, and $\xi_{\sigma\bk}=k^2/(2m_\sigma)-\mu_\sigma$ the kinetic energy measured from the chemical potential.
 The term $2m_r/k^2$ included in eq.\ (\ref{pairPropagator}) ensures the convergence of  $\Pi$ at large momenta, allowing us to express the T-matrix in terms of the physical scattering length.
 It is useful to introduce the energy in the center of mass reference frame, $E_{\rm CM}=\omega-\bP^2/[2(m_\uparrow+m_\downarrow)]+\mu_\uparrow+\mu_\downarrow$.
In the vacuum limit where both chemical potentials vanish, one simply finds $\Pi(\bP,\omega)=-ik_r(m_r/2\pi)$, where $k_r=\sqrt{2m_r E_{\rm CM}}$ is the on-shell relative momentum.
As we see, single channel models containing only a contact four-fermion interaction fail to describe the physics of a narrow resonance, since upon proper renormalization of the coupling constant the physics depends on a single parameter, the scattering length $a$. To make contact with experiments, the latter is usually written in the phenomenological form
 \beq
a=a_{\rm bg}+a_{\rm res}= a_{\rm bg}[1-\Delta B/(B-B_0)].
\label{scatteringLength}
\eeq
Here $a_{\rm bg}$, $\Delta B$ and $B_0$ denote respectively the background scattering length, and the resonance width and center.

A minimal model describing a narrow resonance formally contains two channels (or hyperfine states), termed respectively open and closed. Two particles in the closed channel propagate as a molecule with a magnetic moment $\mu_{c}$ differing from the open channel $\mu_{o}$ by $\delta\mu=\mu_{c}-\mu_{o}>0$.
 The two-channel T-matrix may be written as  \cite{Bruun05,MassignanEfimov08} 
\beq
T(\bP,\omega)=\left[\frac{m_r}{2\pi \tilde{a}(E_{CM})}-\Pi(\bP,\omega)\right]^{-1},
\label{twoChannelTmatrix}
\eeq
where we have introduced the energy-dependent length
\beq
\tilde a(E_{\rm CM})\equiv a_{\rm bg}\left(1-\frac{\Delta B}{B-B_0-E_{\rm CM}/\delta\mu}\right).
\label{aTilde}
\eeq
A detailed derivation of this equation is given in the Appendix.
Comparing with the single-channel T-matrix, we see that the inclusion of the second channel simply results in a replacement of the scattering length $a$ with the energy-dependent quantity $\tilde{a}(E_{CM})$, or equivalently in a magnetic field shift of the resonance center by $E_{CM}/\delta\mu$, towards the BCS (BEC) side for positive (negative) $E_{CM}$.

It is interesting to look closely at the low-energy expansion of the corresponding on-shell scattering amplitude $f(k_r)=-T(\bP,\omega) m_r/2\pi$ . In vacuum, we find
\beq
-f^{-1}(k_r)=a^{-1}+ik_r+R^*k_r^2+O(k_r)^4,
\label{lowEnergyExpansionOfTheScattAmpl}
\eeq
and higher terms may be neglected at the low-energies of interest here.
The range parameter $R^*$ is related to the usual effective range $r_e$ by the relation $R^*=-r_e/2$, and is given by
$R^*=R^*_{\rm res}[1-a_{\rm bg}/a]^2$,
with $R^*_{\rm res}=1/(2m_{r} a_{\rm bg} \Delta B \delta\mu)$.
Since $a\rightarrow+\infty$ as $B\rightarrow B_0^{-}$, one has $a_{\rm bg}\Delta B>0$, and the range parameter $R^*$ is always positive.
Equation (\ref{lowEnergyExpansionOfTheScattAmpl}) reproduces the low-energy physics expected from a generic two-channel model with finite range, in agreement with refs.\ \cite{Petrov04,Gurarie07,Werner09}. In particular, various resonant models such as a square well, a van der Waals potential \cite{Flambaum99}, or a confinement-induced resonance \cite{Massignan06} show the behavior $R^*\propto a^{-2}$ as $a\rightarrow0$.

The pole of the truncated vacuum scattering amplitude eq.~(\ref{lowEnergyExpansionOfTheScattAmpl}) yields the two-body binding energy $E_b=-1/(2m_ra_*^2)$, where the characteristic size of the molecule  $a_*=2R^*/(\sqrt{1+4R^*/a}-1)>0$ interpolates between $a$ for $R^*\ll a$ and $\sqrt{a R^*}$ for $R^*\gg a\gg a_{\rm bg}$; moreover, one has $a^*\rightarrow a_{\rm bg}$ as $a\rightarrow a_{\rm bg}$.

\section{Limiting behaviors}
Given the large number of physical parameters in the problem, it is useful to discuss here some analytic limits of the theory.

To start, we consider the {\it negligible background} regime where $|a_{\rm bg}|\ll|a_{\rm res}|$, see eq.\ (\ref{scatteringLength}). This limit is generally obtained sufficiently close to resonance, where the scattering length diverges. In this limit one finds $a=a_{\rm res}$, $R^*=R^*_{\rm res}$, and $1/\tilde{a}(E_{CM})=a_{\rm res}^{-1}+2R^*m_rE_{CM}$.
To avoid an explicit dependence of our results on $a_{\rm bg}$, we will restrict ourselves in this paper to consider only the {\it negligible background} regime, and in the following we will drop the subscript ``res" to avoid cumbersome notations\footnote{Should instead $a_{\rm bg}$ be positive and large, at the two-body level the T-matrix admits an extra deep bound state with energy $E<-1/(2m_ra_{\rm bg}^2)$, as discussed in ref.\ \cite{MassignanEfimov08}, yielding a finite lifetime for all quasiparticles considered here.}.

We discuss now the condition under which a resonance may be termed broad. For this purpose, in eq.\ (\ref{lowEnergyExpansionOfTheScattAmpl}) we require $|R^*k_r^2|\ll|a^{-1}+ik_r|$. In ultracold Fermi systems, on general grounds one expects bound quasiparticles
 with energy $E\sim E_b$ on the BEC side, and of order $-\epsilon_F$ on the BCS side, while the repulsive polaron will have $E\sim+\epsilon_F$.
Setting $k_r\sim\sqrt{2m_r E}$, to be in the {\it broad resonance} regime one has to satisfy simultaneously the two conditions:
\beq
R^*\ll a\hspace{1cm}{\rm and}\hspace{1cm}R^*\ll k_F^{-1}.
\label{conditionsForUniversality}
\eeq
As a general rule, every s-wave resonance may be considered broad at sufficiently low atom densities and sufficiently close to its center. Given that $R^*\propto g^{-2}$, broad resonances are obtained whenever there is a strong coupling between the open and closed channels.

\section{Quasiparticle equations}
We consider a single impurity perturbing in a negligible way the surrounding Fermi sea. As such, inside the T-matrix we may set $\mu_\downarrow=0$ and $\mu_\uparrow=\epsilon_F$, with $\epsilon_F=k_F^2/2m_\uparrow$ and $k_F=(6\pi^2n_\uparrow)^{1/3}$ the Fermi energy and momentum of the majority atoms.
All polaron properties are then given in terms of the impurity self-energy, which in the one particle-hole approximation (1PHA) and at zero temperature reads \cite{Combescot07}
\beq
\Sigma_{\rm P}(\mathbf{p},E)=\sum_{q<k_{F}}T(\bp+\bq,E+\xi_{q\uparrow}),
\eeq
where $T$ is the T-matrix. The energies of the two polaron branches are given by the two solutions of the equation
\beq
E_\pm=\Re[\Sigma_{\rm P}(\mathbf{p},E_\pm+i0^+)].
\label{polaronEnergy}
\eeq
The polarons residues and effective masses at $\bp=0$ may be obtained as $Z_\pm=[1-\partial_\omega\Re(\Sigma_{\rm P})]^{-1}$ and
\beq
\frac{m^*}{m_\downarrow}=\frac{1}{Z_\pm}\left[1+\frac{\partial \Re(\Sigma_{\rm P})}{\partial(p^2/2m_\downarrow)}\right]^{-1},
\eeq
where the derivatives are taken at the energy of the corresponding quasiparticles. 
The two- and three-body processes leading to the decay of the quasiparticles have been discussed in detail elsewhere~\cite{Bruun10,Massignan11,Kohstall11,Schmidt11}.

The energy $E_{\rm M}$ of a dressed molecule with momentum $\bp$ is given in the 1PHA by the pole of the kernel $K$ of the integral equation \cite{Combescot09,Mora09,Punk09}:
\begin{multline}
\sum_{\bk'} \frac{K_{\bk' \bp \bq}}{E^{(2)}_{\bk \bk' \bp \bq}}
-\sum_{\bq'} \frac{K_{\bk \bp \bq'}}{E^{(1)}_{\bk\bp}}
-\frac{T(\bp,0)}{E^{(1)}_{\bk\bp}}\sum_{\bk' \bq'} \frac{K_{\bk' \bp \bq'}}{E^{(1)}_{\bk'\bp}}\\
+\frac{K_{\bk\bp\bq}}{T(\bq+\bp-\bk,\xi_{q\uparrow}-\xi_{k\uparrow})}
=-\frac{T(\bp,0)}{E^{(1)}_{\bk\bp}},
\label{moleculeEnergy}
\end{multline}
where \mbox{$E^{(2)}_{\bk \bk' \bp \bq}
=E_{\rm M}-\xi_{k\uparrow}-\xi_{k'\uparrow}+\xi_{q\uparrow}-\xi_{(\bk+\bk'-\bp-\bq)\downarrow}$} and $E^{(1)}_{\bk\bp}=E_{\rm M}-\xi_{k\uparrow}-\xi_{(\bk-\bp)\downarrow}$. Holes and particle momenta are restricted to $q<k_F<k,k'$.

 The polaron equation yields the correct result in the limit $|k_Fa|\ll1$, where the attractive/repulsive polaron energies become simply $E_\mp=2\pi a/m_r$. The dressed molecule equation instead becomes exact in the deep BEC regime, where it reduces to $E_{\rm M}=E_b-\epsilon_F+2\pi a_{AD}/m_3$ with $a_{AD}$ the atom-dimer scattering length\cite{Combescot09,Mora09,Punk09,Levinsen11}.
 
To gain further insight on the problem, we consider analytic (variational) upper bounds for the quasiparticle energies at $\bp=0$. For the attractive polaron one has $E_{-}<\min[E_{-}^{\rm Th},E_{0}]$. The Thouless pole for the attractive polaron $E_{-}^{\rm Th}$ is determined by the equation $T^{-1}(k_F,E_{-}^{\rm Th})=0$. The other bound $E_{0}$ is obtained from eq.~(\ref{polaronEnergy}) by constraining the hole momentum $\bq$ to 0. Setting $y_0=-(E_{0}/\epsilon_F)(m_r/m_\uparrow)$, one finds
\beq
\frac{2}{3 y_0}=-\pi(k_F a)^{-1}+\pi k_F R^*y_0+2+2\sqrt{y_0} \arctan{\sqrt{y_0}}.
\eeq
On general grounds, one expects that $E_{\rm P}$ will be closer to  $E_{0}$ in the region where the effective mass of the polaron is positive, while $E_{-}^{\rm Th}$ will be a stricter bound when $m^*<0$ and the molecule has lower energy.
An upper bound for the molecule energy is instead the molecular Thouless pole, given by the solution of $T^{-1}(0,E_{\rm M}^{\rm Th})=0$. Setting $\tilde{y}=-(E_{\rm M}^{\rm Th}/\epsilon_F+1)(m_r/m_\uparrow)$, one finds
\beq
0=-\pi(k_F a)^{-1}+\pi k_F R^*\tilde{y}+2+2\sqrt{\tilde{y}} \arctan{\sqrt{\tilde{y}}}.
\label{molecThoulessPole}
\eeq

\begin{figure}
\onefigure[width=\columnwidth]{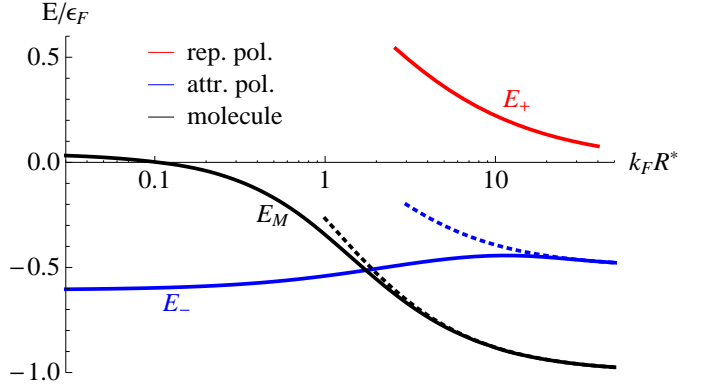}
\caption{Quasiparticle energies at resonance and for equal masses, versus the resonance width $k_FR^*$.
 The dotted lines are the Thouless energies $E_{-}^{\rm Th}$ and $E_{\rm M}^{\rm Th}$.
}
\label{fig:quasiparticlesAtResonanceVskFrStarPetrov_energyAndDecayRate}
\end{figure}
\begin{figure}
\onefigure[width=\columnwidth]{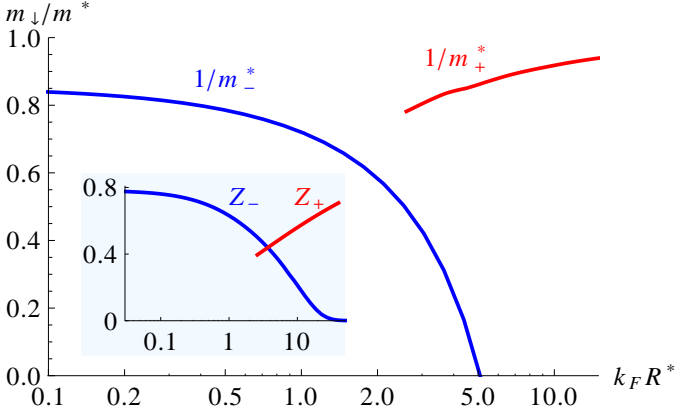}
\caption{Inverse effective mass $m_\downarrow/m^*$ (main fig.) and residue $Z$ (inset) of the two polarons at resonance for $m_\uparrow=m_\downarrow$.}
\label{fig:quasiparticlesAtResonanceVskFrStarPetrov_ZandMeffP}
\end{figure}

\section{Quasiparticle properties at resonance}
We focus here on the case $m_\uparrow=m_\downarrow$, and at $(k_Fa)^{-1}=0$.
In fig.~\ref{fig:quasiparticlesAtResonanceVskFrStarPetrov_energyAndDecayRate} we show how the energies of the $\bp=0$ polarons and dressed molecule  are affected by the narrow character of the Feshbach resonance, parametrized by $k_FR^*$.
In the broad resonance limit $k_FR^*\ll 1$ the attractive polaron is the ground state of the system with an energy $E_{-}=-0.61\epsilon_F$, while the dressed molecule experiences strong repulsive interactions with the gas, and lies at energy $E_M=0.05\epsilon_F$, more than one Fermi energy above its ground state in the absence of interactions. As the resonance becomes narrower, the two quasiparticle branches come closer, touch at $k_FR^*_x=1.7$, and finally the molecule becomes the ground state of the mixture. In the asymptotic limit $k_FR^*\gg 1$, particle-hole dressing becomes negligible and the two energies approach the analytic upper bounds given by the respective Thouless poles, $E_{-}^{\rm Th}$ and $E_{\rm M}^{\rm Th}$. In the vicinity of the resonance, from eq.\ (\ref{molecThoulessPole}) one finds
\beq
E_{\rm M}^{\rm Th}=-{\epsilon_F}\left[1-\frac{m_\uparrow}{m_r}\frac{2-\pi(k_Fa)^{-1}}{2+\pi k_F R^*}\right].
\label{analyticMolecThoulessPoleNarrow}
\eeq
Since $E_{\rm M}^{\rm Th}>E_{\rm M}\geq E_b-\epsilon_F$, and $E_b=0$ at resonance, eq.\ (\ref{analyticMolecThoulessPoleNarrow}) gives the first terms of a perturbative expansion of the ground state energy $E_{M}$ for $|k_Fa|^{-1}\ll 1$ and $k_FR^*\gg 1$.

The repulsive polaron is not a good quasiparticle at resonance in the broad resonance limit, as its decay rate is too large when compared to its energy \cite{Massignan11,Kohstall11,Schmidt11}. Nonetheless, a well-defined repulsive polaron appears at unitarity for $k_FR^*\gg 1$, as in this limit its decay rate becomes negligible when compared to its energy $E_{+}$.

The inverse effective mass $1/m^*$ and residue $Z$ of the polarons are shown in fig.\ \ref{fig:quasiparticlesAtResonanceVskFrStarPetrov_ZandMeffP}. Leaving the broad resonance regime, both quantities decrease for the attractive polaron, and increase for the repulsive one. In particular, beyond the critical width $k_FR^*_x$ the effective mass diverges and becomes negative, a usual signature of the ground state transition from a polaron to a molecule \cite{Prokofev08}. Correspondingly, increasing spectral weight is continuously transferred from the attractive to the repulsive branch.

\begin{figure}
\onefigure[width=\columnwidth]{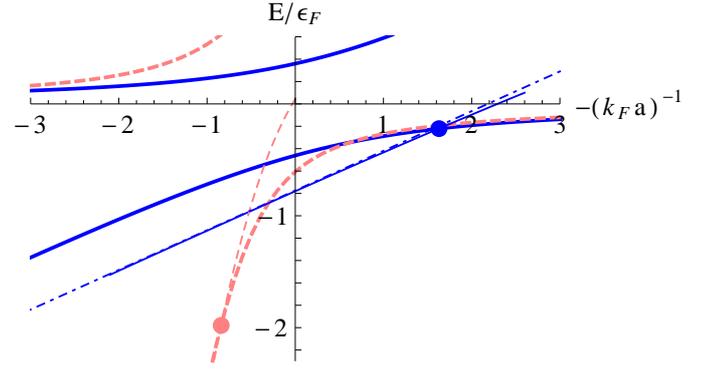}
\caption{Quasiparticle energies across the resonance for $m_\uparrow=m_\downarrow$. Polarons (molecules) are depicted by thick (thin) lines. Continuous blue lines correspond to $k_FR^*=5$, while dashed pink ones are for $k_FR^*=0$. The polaron/molecule crossings are marked by dots. The dot-dashed blue line is the analytic result for the molecule energy with $k_FR^*=5$ from eq.\ (\ref{analyticMolecThoulessPoleNarrow}).}
\label{fig:plotEnergyCrossover}
\end{figure}

\begin{figure}
\onefigure[width=\columnwidth]{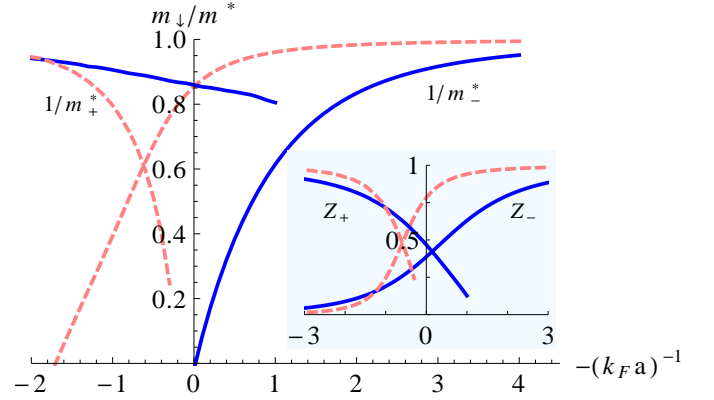}
\caption{Inverse effective mass (main figure) and residue (inset) of the two polarons for $m_\uparrow=m_\downarrow$. Lines as in fig.~\ref{fig:plotEnergyCrossover}.}
\label{fig:plotInvEffMassAndResidueCrossover}
\end{figure}

\begin{figure}
\onefigure[width=\columnwidth]{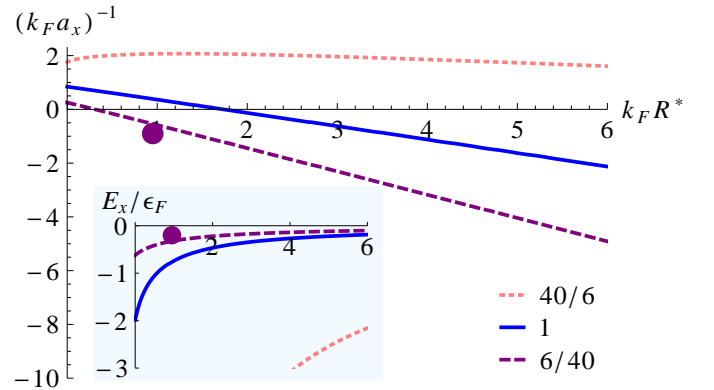}
\caption{Critical interaction strength of the polaron/molecule crossing as a function of the resonance width for various mass ratios $m_\uparrow/m_\downarrow$. Above (below) the line the ground state is a molecule (polaron).
 Inset: energy $E_x$ of the excitations at the crossing.   The dots mark the interaction strength and energy of the crossing as located in the K-Li mixture of ref.~\cite{Kohstall11}.}
\label{fig:polMolCrossing}
\end{figure}

\section{Quasiparticle properties through the crossover}
The features displayed by the quasiparticles at resonance are actually quite general, and reproduced throughout the BEC-BCS crossover and for a large range of mass ratios. Our theoretical results for potassium impurities in a lithium gas with $k_FR^*\approx 1$ were presented in ref.~\cite{Kohstall11}, and agreed with remarkable accuracy with experimental observations.
Here we focus on the equal masses case, and compare in figs.~\ref{fig:plotEnergyCrossover}-\ref{fig:plotInvEffMassAndResidueCrossover} the polaron properties for $k_FR^*=5$ to known results from the broad case.
 On the BEC side, the attractive polaron and the dressed molecule follow the two-body binding energy $E_b$, and appear less bound than in the broad case. On the BCS side, the repulsive interaction experienced by the dressed molecule is reduced, yielding a dressed molecule increasingly bound deep in the BCS regime, and a shift of the polaron/molecule crossing in the same direction. Remarkably, eq.\ (\ref{analyticMolecThoulessPoleNarrow}) gives a very good approximation to $E_{\rm M}$ close to resonance even for a moderately narrow resonance with $k_F R^*=5$. The energy-dependence of the T-matrix is important for narrow resonances, as noted below eq.~(\ref{scatteringLength}), and yields sensible shifts of all other quasiparticle properties, such as the polarons' inverse effective mass and residue shown in fig.\ \ref{fig:plotInvEffMassAndResidueCrossover}.

 The critical interaction strength $(k_Fa_x)^{-1}$ at which the polaron and molecule energies are equal is plotted in fig.~\ref{fig:polMolCrossing} for various mass ratios. For every interaction strength, the dressed molecule becomes the ground state of the mixture at sufficiently large $k_FR^*$. For $m_\uparrow/m_\downarrow=6/40$ and $k_F R^*=0.95$, we find $-(k_Fa_x)^{-1}=0.6$ in good agreement with the value obtained in the imbalanced K-Li mixture of ref.~\cite{Kohstall11} (dots in fig.~\ref{fig:polMolCrossing}). The residual small discrepancy may partially be ascribed to finite experimental resolution, but especially to non-zero temperature effects:  $E_{-}$ and $E_{M}$ intersect at a very small angle, and a minute shift of the energy levels can yield a noticeable shift of the crossing. The inset of fig.~(\ref{fig:polMolCrossing}) shows the energy $E_x$ of the two excitations at the crossing.
In the narrow limit $k_FR^*\gg 1$, we find $(k_Fa_x)^{-1}\sim-(k_FR^*) m_r/m_\uparrow$ and  $E_x\propto (k_FR^*)^{-1}$, as one recovers analytically from the corresponding asymptotics $E_{0}\sim E_{\rm M}^{\rm Th}$ (as in this limit $E_{-}\lesssim E_{0}<0<E_{\rm -}^{\rm Th})$.

\section{Tan's relation for the size of the dressing cloud}
The energy density of a Fermi sea containing a few impurities ($n_\downarrow\ll n_\uparrow$) is given by \cite{Punk09}
\begin{equation}
\varepsilon=(3/5)\epsilon_F n_\uparrow+E_\downarrow n_\downarrow,
\label{energyDensity}
\end{equation}
where $E_\downarrow=\min[E_{-},E_{M}]$ is the ground state energy of a single dressed impurity. The self-consistent equation $E_{-}=\Sigma_{\rm P}(E_{-})$  at $\bp=0$ gives the polaron energy, and a similar one may in principle be written for the dressed molecule energy in terms of a self-energy $\Sigma_{\rm M}$. We restrict ourselves here to the broad resonance case, where both self-energies are simply functions of the energy $\omega$ and of the scattering length $a$. The derivative of eq.~(\ref{energyDensity}) with respect to $v=1/a$ yields Tan's contact density $\mathcal{C}$ \cite{Tan08, Werner10}
\begin{equation}
-\frac{\mathcal{C}}{8\pi m_r}=\frac{{\rm d} \varepsilon}{{\rm d} v}
=n_\downarrow \left(\frac{\partial \Sigma}{\partial v}+\frac{\partial \Sigma}{\partial \omega}\frac{\partial E_\downarrow}{\partial v}\right)
=n_\downarrow Z \frac{\partial \Sigma}{\partial v},
\label{TanContact}
\end{equation}
where $\Sigma$ and $Z$ are the ground state (polaron/molecule) self-energy and residue, and derivatives are evaluated at the quasiparticle energy. The number $\Delta N$ of particles in the dressing cloud of the impurity was shown to be \cite{Massignan11}
\beq
\Delta N=-\frac{\partial \Sigma}{\partial\epsilon_F}=-\frac{E_\downarrow}{\epsilon_F}+\frac{Z}{2 a \epsilon_F}\frac{\partial \Sigma}{\partial v}.
\label{DeltaN}
\eeq
Combining the latter two equations, we find
\beq
\frac{\mathcal{C}}{16\pi m_r a n_\downarrow}=-E_\downarrow-\Delta N \epsilon_F.
\label{CvsDeltaN}
\eeq

 Since the contact remains finite through the crossover, at the unitary point of a broad resonance we find \mbox{$\Delta N=-E_\downarrow/\epsilon_F$}.

It can be shown that equation (\ref{CvsDeltaN}) is actually a limiting case of Tan's pressure relation for the impurity problem \cite{TanPrivate}. Indeed, the pressure of an ideal homogeneous Fermi gas is $\mathcal{P}_0=2 \epsilon_Fn_\uparrow/5$, and its change $\Delta \mathcal{P}$ with a change in number of particles $\Delta n_\uparrow$ is $\Delta \mathcal{P}= \mathcal{P}- \mathcal{P}_0=2 \epsilon_F (\Delta n_\uparrow)/3$. Since $\Delta n_\uparrow=-n_\downarrow \Delta N$ \cite{Massignan11}, and $\mathcal{C}_0=0$ in absence of impurities, eq.~(\ref{CvsDeltaN}) may be rewritten in the form of Tan's pressure relation \cite{Tan08}, 
\begin{equation}
\frac{\mathcal{C}}{24 \pi m_r a}=\mathcal{P}- \frac{2}{3}\varepsilon,
\label{pressureRelation}
\end{equation}
where $\varepsilon$ is the energy density of the ideal gas upon addition of a few impurities, as given in eq.\ (\ref{energyDensity}).

\section{Conclusions}
In this paper we investigated the properties of impurities in a Fermi sea via a many-body two-channel model, the minimal formalism which captures quantitatively the physics of a Feshbach resonance with arbitrary width. We calculated the properties of the relevant quasiparticles, showing how the critical interaction strength for the polaron/molecule crossing evolves as a function of resonance width and mass ratio. Moreover, the ground state energy close to resonance has been proved to become perturbative in the small parameter $1/(k_FR^*)$.

 For broad resonances, we have also derived an alternative Tan's pressure relation, linking the impurity energy $E_\downarrow$ to the number of atoms in its dressing cloud $\Delta N$.

 Results presented here are important to understand on-going studies on hetero-nuclear ultracold mixtures, specially in view of the recent demonstration that narrow resonances provide an ideal framework for the realization of an elusive state, the strongly-repulsive Fermi gas. We hope that our results will stimulate further investigations of interacting quantum mixtures.

\textit{Additional remark:} shortly after the submission of this manuscript, related results appeared in \cite{QiTrefzger}.

\section{Appendix: T-matrix for narrow resonances}

\begin{figure}
\onefigure[width=\columnwidth]{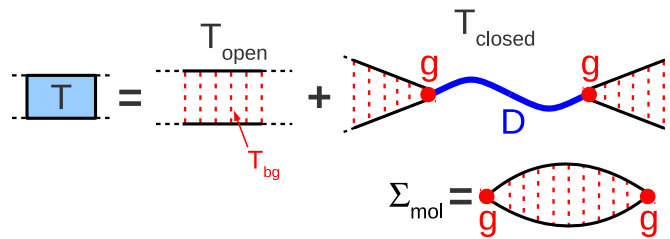}
\caption{Diagrammatic structure of the two-channel T-matrix. Straight and wavy lines indicate respectively propagation in the open and closed channels.}
\label{diagrams}
\end{figure}

We write the two-channel T-matrix as
\beq
T(\bP,\omega)=T_{\rm open}(\bP,\omega)+T_{\rm closed}(\bP,\omega).
\label{Tinitial}
\eeq
The corresponding Feynman diagrams are sketched in fig.\ \ref{diagrams}, and we will discuss them separately below. 

$T_{\rm open}$ describes particles propagating only in the open channel, undergoing repeated scattering processes parametrized by the background scattering length $a_{\rm bg}$. In the ladder approximation, the infinite series of scattering events may be resummed to give
\beq
T_{\rm open}^{-1}(\bP,\omega)=T_{\rm bg}^{-1}-\Pi(\bP,\omega).
\label{Topen}
\eeq
Here $T_{\rm bg}=2\pi a_{\rm bg}/m_{r}$, and  $\Pi(\bP,\omega)$ is the renormalized pair propagator given in eq.\ (\ref{pairPropagator}).

$T_{\rm closed}$ describes instead scattering events that couple two particles in the open channel to the closed channel. The closed channel molecule experiences an energy shift $\delta\mu(B-B_0)$ with respect to two particles in the open channel, and its propagator reads
\beq
D(\bP,\omega)=[E_{\rm CM}-\delta\mu(B-B_0)-\Sigma_{\rm mol}(\bP,\omega)]^{-1}.
\label{moleculePropagator}
\eeq
We have assumed here that medium effects arising from the non-zero population of the closed channel are negligible at all times. This assumption is surely justified when considering few impurities in a Fermi sea, as there may be at most as many closed-channel molecules as impurities.
The two atoms may perform repeated scattering events in the open channel before ending up in the closed channel, a process overall described by the ``dressed vertex" $V(\bP,\omega)=g[1-T_{\rm bg}\Pi(\bP,\omega)]^{-1}$, with $ g=\sqrt{T_{\rm bg} \Delta B \delta \mu}$  \cite{Bruun04} the renormalized coupling strength between two atoms in the open channel and a molecule in the closed channel\footnote{
As we limit ourselves to Fermi-Fermi mixtures with mass ratios $m_\uparrow/m_\downarrow$ smaller than 13.6, the results are independent of short-distance physics and a momentum-independent interaction may safely be taken.
Effects arising from a non-zero range of the interaction vertices are instead crucial for systems exhibiting the Efimov effect, such as bosonic gases, or Fermi-Fermi mixtures with mass ratios $m_\uparrow/m_\downarrow>13.6$. Their inclusion is discussed e.g.\ in ref.\ \cite{Bruun05,Werner09}.}.

 The self-energy $\Sigma_{\rm mol}$ describes a molecule coupling to two atoms [$g$], which propagate in the Fermi sea [$\Pi$] and recombine into a molecule through a dressed vertex [$V$]:
\beq
\Sigma_{\rm mol}(\bP,\omega)=g \, \Pi(\bP,\omega) \, V(\bP,\omega),
\label{moleculeSelfEnergy}
\eeq
 Only one of the two vertices in eq.\ (\ref{moleculeSelfEnergy}) has been  dressed to avoid double-counting.
The contributions coming from the coupling to the closed channel finally yield
\beq
T_{\rm closed}(\bP,\omega)=V(\bP,\omega)D(\bP,\omega)V(\bP,\omega).
\label{Tclosed}
\eeq

Inserting eqs.\ (\ref{Topen}) and (\ref{Tclosed}) into (\ref{Tinitial}), one finds the compact expression for the T-matrix given in eq.\ (\ref{twoChannelTmatrix}).

\acknowledgments
Insightful discussion with C. Kohstall, R. Grimm, M. Lewenstein, S. Tan, M. Zaccanti, and in particular  G. M. Bruun and J. Levinsen are gratefully acknowledged. We wish to thank the Aspen Center for Physics, where part of this work has been realized. This research has been funded through ERC Advanced Grant QUAGATUA, and Spanish MEC project TOQATA.

\end{document}